\documentclass[reqno, 11pt]{amsart}
\usepackage{amsmath,amssymb,color}
\usepackage[mathscr]{euscript}
\usepackage{stmaryrd}
\usepackage{graphicx}
\usepackage[pdftex,breaklinks,colorlinks,
citecolor=blue,
urlcolor=blue,
pdftitle={On the hydrodynamics of active matter models on a lattice},
pdfauthor={C. Erignoux}]{hyperref}
\usepackage{amsfonts, amscd, epsfig, amsmath, amssymb,enumerate}
\usepackage{graphicx}
\usepackage{color}
\usepackage{mathrsfs}
\usepackage[margin=3cm]{geometry}
\usepackage{tikz}
\usetikzlibrary{backgrounds}
\usetikzlibrary{patterns,fadings}
\usetikzlibrary{arrows,decorations.pathmorphing}
\usetikzlibrary{decorations.pathreplacing}
\usetikzlibrary{decorations}
\usetikzlibrary{calc}
\usetikzlibrary{shapes.misc}
\definecolor{light-gray}{gray}{0.95}

\usepackage{float}
\def\centerarc[#1](#2)(#3:#4:#5){\draw[#1] ($(#2)+({#5*cos(#3)},{#5*sin(#3)})$) arc (#3:#4:#5);}

\allowdisplaybreaks 

\makeatletter
\@addtoreset{equation}{section}
\makeatother

\newtheorem{theorem}{Theorem}[section]

\newtheorem{remark}[theorem]{Remark}

\newcommand{\bb}[1]{{\mathbb #1}}

\newcounter{as}[section]

\newcommand{\R}{{\mathbb R}}

\newcommand{\N}{{\mathbb N}}

\newcommand{\T}{{\bb{T}}}
\newcommand{\E}{{\mathbb{E}}}
\newcommand{\Prob}{{\mathbb{P}}}

\newcommand{\pa}[1]{\left(#1 \right)}

\newcommand{\cro}[1]{\left[#1\right]}

\DeclareMathOperator{\Pe}{Pe}

\definecolor{dkgreen}{rgb}{0,0.6,0}
\definecolor{gray}{rgb}{0.5,0.5,0.5}
\definecolor{header}{gray}{0.3}

\title[On the hydrodynamics of active matter models on a lattice]{On the hydrodynamics of active matter models on a lattice}
\author{C. Erignoux}
\email{clement.erignoux@inria.fr}
\address{Equipe PARADYSE, Bureau B211
Centre INRIA Lille Nord-Europe
Park Plaza, Parc scientifique de la Haute-Borne, 40 Avenue Halley B\^atiment B, 59650 Villeneuve-d'Ascq
France}
\date{\today}

\begin{document}
\begin{abstract}
Active matter has been widely studied in recent years because of its rich phenomenology, whose mathematical understanding is still partial. We present some results, based on \cite{Erignoux,KEBT} linking microscopic lattice gases to their macroscopic limit, and explore how the mathematical state of the art allows to derive from various types of microscopic dynamics their hydrodynamic limit. We present some of the crucial aspects of this theory when applied to weakly asymmetric active models. We comment on  the specific challenges one should consider when designing an active lattice gas, and in particular underline mathematical and phenomenological differences   between gradient and non-gradient models. Our purpose is to provide the physics community, as well as member of the mathematical community not specialized in the mathematical derivation of scaling limits of lattice gases, some key elements in defining microscopic models and deriving their hydrodynamic limit. 
\end{abstract}
\pagenumbering{Alph}
\maketitle
\thispagestyle{empty}
\pagenumbering{arabic}


\section*{Introduction}


Active matter has been the subject of intense scrutiny in recent decades, across different scientific communities. From the perspective of \emph{individual-based} (or \emph{agent-based}) models, active matter is composed of many individuals interacting with their surrounding, and consuming energy individually to self-propel. Active matter models have been used by the physics community to simulate various types of biological, physical and chemical behavior, from animal flocking \cite{animal}, to bacterial motion \cite{bacteria}, to metallic rod's spontaneous spatial organization \cite{rods}. Interest for active matter models arose from Vicsek and coauthor's seminal work \cite{Vicsek}, where  a phase transition phenomenon was uncovered numerically for a particle system where individuals locally align their velocities up to a small stochastic noise. Since then, alignment phase transition was identified as an ubiquitous phenomenon \cite{Alignment0, Alignment1, Alignment2, Alignment3}, spanning many types of models with an alignment mechanism between particle's velocities. 

\medskip


Active matter can also form spontaneous condensates, when particle's velocities decrease in crowded regions (see \cite{CT2015} and references therein for an exhaustive review, \cite{SSC18,SFC20} for recent developments). This phenomenon, known in the physics community as \emph{Motility Induced Phase Separation}, or MIPS, results from the positive feedback between clustering and slow down of particles: the more particles aggregate, the more they slow down, and the more they aggregate. It translates as a phase separation between a dense "liquid" aggregate, and a low density "gaseous" phase. This type of behavior does not occur in passive systems, in which aggregation would be followed by diffusion, and thus spreading out of the aggregate. 


From a mathematical point of view, the efforts of the community have focused on \emph{mean-field} and \emph{locally mean-field} models  of active matter (see for example \cite{DM2007, Frouvelle2011, FL2012}), in which the interaction of each particle with its environment is averaged out over its (small) macroscopic neighborhood. From a physics standpoint, this is equivalent to replacing the microscopic observables of the system by their average field. The  mean-field assumption simplifies a number of difficulties concerning out-of equilibrium active models, and can allow for explicit derivation of hydrodynamic limits in the continuum (see \cite{DM2007} for a model close to Vicsek's original model), as well as the fluctuations around it \cite{DY2010}. 

Unfortunately, it is in general not clear that the mean-field assumption is a reasonable one, and from that perspective it remains fundamental to be able to derive mathematical results from models with local interactions. A fundamental flaw in the mean-field approach is that it can fail to capture parts of the behavior of the model studied. As an example, although Viscek model's  phase transition is now understood to be first order \cite{GC2004}, its mean-field iteration, which have been the subject of intense scrutiny from the analysis community,  has a continuous phase transition \cite{DB14}: because of its sensitivity to noise \cite{MCN20}, the mean-field model fails to capture the first order scenario of the Vicsek model.

\medskip


For this reason, in order to further the mathematical understanding of active matter, it is worthwhile to look past mean-field or local mean-field--type interactions, and derive mathematical results on the phenomenology of active matter starting from individual based models where particles interact purely on a microscopic scale.  An important tool to achieve this program is the theory of hydrodynamic limits, which was broadly used by both mathematical and physics communities to characterize the large scale behavior of microscopic particle systems. 


One of the most fundamental and widely studied class of models to which this theory can be applied is the class of \emph{lattice gases} (see the monograph \cite{KL} and references therein), in which particles evolve stochastically on a discrete lattice. Letting the mesh size $\varepsilon=1/N$ go to $0$, under proper rescaling of time and space, allows for explicit and rigorous derivation of the scaling limit of the system under fairly broad  assumptions on the particle's dynamics.


\medskip

 The mathematical theory of hydrodynamic limits, and scaling limits (e.g. study of fluctuations and large deviations) in general, is well established for lattice gases and has resulted in significant achievements by the mathematical community to tackle a range of problems in non-equilibrium statistical mechanics. It is, however, a rather technical topic, and even though many leading researchers in the field of particle system's scaling limits have a split background between mathematics and theoretical physics, the general mathematical formalism and techniques remains fairly inaccessible to the physics community at large.

\medskip

The purpose of this note is twofold: first, we present a few key concepts that are of fundamental importance in the mathematical study of lattice gases. Our purpose will be to give a reader, even unfamiliar with the mathematical formalism, a sense of both the aspects of microscopic models which can result in significant technical difficulties from a hydrodynamic limit standpoint, and the key ideas behind the mathematical theory of hydrodynamic limits. Second, we address some specific challenges when applying this theory to active matter models, in particular in the context of trying to obtain a  phenomenological understanding of the model's macroscopic behavior. We will, to do so, be  using elements of \cite{Erignoux} to illustrate (extreme) mathematical challenges, as well as \cite{KEBT} to illustrate how these techniques can result in tangible information on phase separation phenomena in active matter at a reasonable mathematical cost. We will also briefly address the topic of fluctuating hydrodynamics to emphasize how lower order terms can also be obtained in a mathematically rigorous way. Of course, our purpose here is not to give a precise overview of the mathematical state of the art (for that purpose, see for example \cite{KL}), but rather to illustrate it using natural choices of lattice gases to model active matter, namely \emph{active exclusion processes} and active \emph{zero-range processes}. 

\medskip

We emphasize that we mainly illustrate two of the main tools to prove hydrodynamic limits, namely Guo, Papanicolaou and Varadhan's \emph{entropy method} \cite{GPV1988} and Varadhan's \emph{non-gradient method} \cite{Varadhan1994b}. Other techniques, like Yau's \emph{relative entropy method}  \cite{Yau1991} or duality-based methods go beyond the scope of this note.

\medskip

This article is organized as follows: in Section \ref{sec:WAEPZR}, we introduce a mathematical formalism for active lattice gases, with a specific emphasis on two of the most widely studied  types of models, namely \emph{exclusion processes} and \emph{zero-range processes}. Section \ref{sec:hydrograd} is devoted to illustrate how the \emph{entropy method} \cite{GPV1988} can be applied to the  simple active lattice gases studied in \cite{KEBT} in order to derive their hydrodynamic limit in a mathematically rigorous way, and in turn how exact phase diagram both for alignment phase transition and Motility-Induced Phase Separation can be obtained. We finish the section by explaining the key concept of \emph{local equilibrium}, and how one can deduce from it the shape of the hydrodynamic limit. Section \ref{sec:NG} is devoted to non-gradient models, which are mathematically more challenging. Based on a simpler version of the model studied in \cite{Erignoux}, we explain there the definition of non-gradient models, and how this characteristic of the microscopic model impacts the derivation of its hydrodynamic limit, by Varadhan's non-gradient tools \cite{Varadhan1994b}. We discuss in Section \ref{ref:CGscale} the microscopic, mesoscopic and macroscopic  coarse-graining scales which lie at the center of most modern tools used to derive hydrodynamic limit. Finally, we discuss in Section \ref{sec:extensions} natural generalizations of the models and techniques presented above.

\section{Weakly asymmetric exclusion and zero-range processes}
\label{sec:WAEPZR}

Consider the $1$-dimensional  discrete lattice $\T_N=\{1,\dots, N\}$, with periodic boundary conditions. This lattice is seen as a discretization of the (macroscopic) continuous segment $[0,1]$, so that the mesh size is $1/N$. Each site $x$ of the lattice is, at any given time $t$, in a state $\eta_x(t)\in \chi$, where $\chi$ represents each site's state-space. In the context of active lattice gases, each site will be occupied by a number of particles, each with a given velocity. For clarity of exposition, we will focus for now on cases where only two "velocities" $\pm$ are possible for each particles, we will address later on more general cases (see section \ref{sec:continuum}). We  consider   two of the most widely studied types of lattice gases,  namely 
\begin{itemize}
\item \emph{exclusion processes} in which at most one particle occupies each site: $\chi=\{0,+1, -1\}$, where $\eta_x=+1$ (resp. $-1$)  if site $x$ is occupied by a particle of type $+$ (resp. $-$), and $\eta_x=0$ if site $x$ is empty.
\item \emph{Zero-range processes},  in which each site can contain an arbitrarily large number of particles: $\chi=\{(n^+, n^-)\in \N^2\}$, $n^+$, resp $n^-$ representing the number of particles of type $+$ (resp. type $-$) present at the site. For zero-range processes, the evolution of a particle at site $x$ only depends on the local state $\eta_x$ of the configuration on its site.
\end{itemize}
A \emph{configuration} for the system is given by a function $\eta=(\eta_x)_{x\in \T_N}\in \chi^{\T_N}$.

\medskip

From the standpoint of active lattice gases, three dynamical components should naturally appear throughout the evolution of the system :
\begin{itemize}
\item Symmetric, nearest-neighbor particle jumps: although Vicsek's model, for example, does not contain a diffusive component in particle's motion, from a mathematical standpoint, allowing particles to jump symmetrically throughout the system is crucial. Indeed, as will be emphasized later on, fast diffusion allows the system to mix quickly, and maintain a state of \emph{local equilibrium} (cf. Section \ref{sec:LE}). In the context of this note, these symmetric particle jumps will occur at a rate $DN^2$, $D$ being the diffusive constant of the dynamics.

\item Asymmetric nearest-neighbor particle jumps depending on the "velocity"  $\pm$ of the particle, representing the active nature of the lattice gas. Under this part of the dynamics, particles of type $+$ jump to the right, whereas particles of type $-$ jump to the left. We will focus here on \emph{weakly asymmetric processes, } in which these asymmetric jumps occur at a rate $\lambda N$, the parameter  $\lambda\in \R^+$ tuning the weak asymmetry.

\item A flipping mechanism of Glauber type, allowing particle to change their velocity $ \pm$, depending or not on the local state of the system. In other words, a particle $\pm$ at site $x$ changes type and becomes $\mp$ at a rate $c_x(\eta)$. This rate does not scale with $N$.
\end{itemize}
Let us comment further on the scalings of the three parts of the dynamics. Since the symmetric jumps are not affected by the particle's type, the first part of the dynamics, scaling as $N^2$, roughly amounts to a a symmetric random walk, which lets  particles travel a distance $O(N)$ on the lattice in a time of order one. The asymmetric jump always occur in the same direction as long as the particle does not change type, therefore a particle will also travel with asymmetric jumps a distance of order $O(N)$ in a time of order one, provided it does not update its type. Finally, the Glauber dynamics occurs at rate of order $1$, which means that a particle will typically travel macroscopic distances (of order $N$) between two Glauber updates. Because of this, the three components of the dynamics will appear on equal footing at the hydrodynamic limit. Throughout, all processes evolve on macroscopic time scales (which means that the microscopic processes are all accelerated by the relevant scalings in $N$ introduced above), and the variable $x$ represents a discrete space variable, whereas $u\simeq x/N$ represents the continuous space variable.

\medskip

It is noteworthy that the hydrodynamic limit theory that is briefly outlined below does not a priori require  these exact scaling : one could for example consider weaker symmetric motion on a scale $N^\delta\ll N^2$, without necessarily losing the mixing properties crucial to the derivation of the hydrodynamic limit. However, doing so removes the diffusive part of the macroscopic equation, and one can no longer perform the linear stability analysis yielding exact phase diagrams for phase separation \cite{KEBT}.


\section{Hydrodynamics of active lattice gases}
\label{sec:hydrograd}

We now describe two specific models, considered in \cite{KEBT}, to illustrate key ideas and steps in the derivation of hydrodynamic limits, which in turn allows to obtain exact phase diagrams for MIPS and Vicsek--type alignment phase transition. We will consider in the next section a third one, which is a simplified version of the model considered in \cite{Erignoux}, and will serve to illustrate the importance of gradient lattice gases. 
\subsection{A simple MIPS model}
\label{sec:MIPSmodel}
The first model considered in \cite{KEBT} is a one-dimensional \emph{exclusion process}, that is $\eta_x\in\{0,+1, -1\}$. As outlined in the previous section, there are three components to the dynamics:
\begin{enumerate}[i)]
\item Two neighboring sites $x$, $x+1$ exchange their content $\eta_x$, $\eta_{x+1} $ at rate $DN^2$.
\item A particle $\pm$ at site $x$ jumps at site $x\pm1$ at rate $\lambda N$ \emph{if it is empty} (this last constraint is called \emph{exclusion rule}).
\item A particle changes type at constant rate $\gamma$.
\end{enumerate}

We fix two initial profiles $\rho_0^+$, $\rho_0^-:[0,1]\to [0,1]$ such that $\rho_0:=\rho_0^++\rho_0^-\leq 1$, and consider an initial state $\eta(0) $ for the system defined by 
\begin{equation}
\label{eq:initstate}
\eta_x(0)=\begin{cases}
1 &\mbox{ w.p. }\rho_0^+(x/N),\\
-1 &\mbox{ w.p. }\rho_0^-(x/N),\\
0 &\mbox{ w.p. }1-\rho_0^+(x/N)-\rho_0^-(x/N)\\
\end{cases}
\end{equation}
independently for each $x\in \T_N$.

\medskip

For convenience, given a configuration $\eta=(\eta_x)_{x=1\dots, n}$, we define 
\begin{equation}
\label{eq:sigma}
\sigma^\pm_x={\bf 1}_{\{\eta_x=\pm 1\}} \quad \mbox{ and }\quad \sigma_x=|\eta_x|=\sigma^+_x+\sigma_x^-,
\end{equation}
the latter representing the number of particles at site $x$. The initial state defined by \eqref{eq:initstate}, together with the dynamics i)-iii) above defines a  continuous-time Markov process $(\sigma^\pm(t))_{t\geq 0}$ whose macroscopic limit $N \to\infty$ can be defined in several ways. The most mathematically satisfying one involves a space of measures on $[0,1]$, however, it is also the most burdensome. Instead, we therefore settle here for more intuitive definitions. The first way to look at the macroscopic limit of the process involves the discrete density fields  $\rho^{\pm,N}_x(t)=\E(\sigma^\pm_x(t))$. The process $(\sigma^\pm(t))_{t\geq 0}$ is then characterized at the macroscopic scale by its \emph{hydrodynamic limit} $\rho^\pm(t,u)$ which is the scaling limit of the density fields, defined for any $u\in[0,1]$ as
\begin{equation}
\label{eq:EDF} 
\rho^\pm(t,u)=\lim_{N\to\infty}\rho^{\pm,N}_{\lfloor Nu\rfloor}(t).
\end{equation}

A  second way to look at the macroscopic field is as a \emph{coarse-grained} limit of the process: fix $0<\delta<1$, we can also define 
\begin{equation}
\label{eq:CGF}
\rho^{\pm}(t,u)=\lim_{N\to\infty}\frac{1}{2N^\delta+1}\sum_{|x-uN|\leq N^\delta}\sigma^\pm_x(t).
\end{equation}
The fact that these two limits coincide, and that the second does not depend on $\delta\in ]0,1[$, is of course not obvious, and is a consequence of \emph{local equilibrium}, as developed further below in Section \ref{sec:LE}.

\medskip

Many mathematical tools have been developed in the last decades to derive scaling limits of lattice gases. In the case of the active lattice gas described above, the celebrated \emph{entropy method} developed by Guo, Papanicolaou and Varadhan \cite{GPV1988} yields the following result which characterizes the \emph{hydrodynamic limit} of the process.
\begin{theorem}
\label{thm:hydro1}
Recalling \eqref{eq:CGF}, define the density field $\rho=\rho^++\rho^-$, as well as the magnetization field $m=\rho^+-\rho^-$.  The coarse-grained fields $\rho,\;m$ are solution to the coupled equations 
\begin{equation}
\label{eq:HDLMIPS}
\begin{split}
\partial_t \rho&=D\partial_{uu} \rho-\lambda \partial_u m(1-\rho),\\
\partial_t m&=D\partial_{uu} m-\lambda \partial_u \rho(1-\rho)-2\gamma m,
 \end{split}
\end{equation}
with initial data $\rho(0,\cdot)=\rho_0^++\rho_0^-$, $m(0,\cdot)=\rho_0^+-\rho_0^-$.
\end{theorem}

\begin{figure}
\includegraphics[width=10cm]{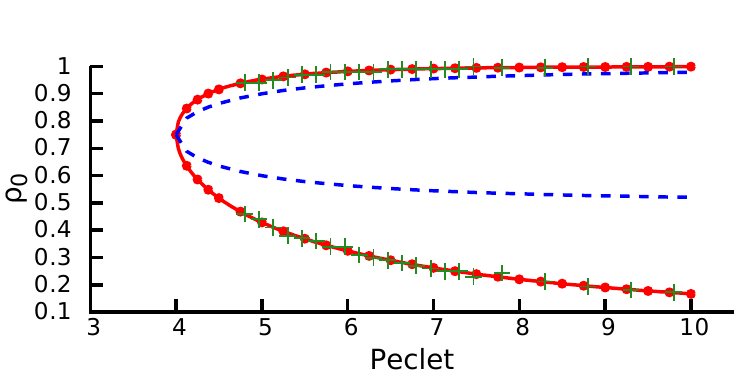}
\caption{From \cite{KEBT}: the phase diagram for the linear stability of the MIPS model presented in Section \ref{sec:MIPSmodel}. The spinodal curve is represented in blue. Outside the spinodal region, the uniform profile $\rho\equiv \rho_0$ is linearly stable. Inside, it is unstable, thus leading to two fully separated phases, a \emph{liquid phase} whose density is given by the top red curve, and a low-density \emph {gaseous phase} whose density is given by the bottom red curve. Comparison with simulations of the microscopic system (green crosses) show perfect agreement with the analytical densities.}
\label{fig:MIPS}
\end{figure}

Because of the exclusion rule on the asymmetric jumps, particles slow down in crowded regions, and this model exhibits Motility-Induced Phase Separation (MIPS). The exact derivation of the hydrodynamic limit allows to obtain an exact phase diagram for MIPS, represented in Figure \ref{fig:MIPS}. 

\medskip

To do so, consider the two-dimensional equivalent of the model above, where there are still two types of particles $\pm$, jumping to $x\pm e_1$ at an extra rate $\lambda N$. we reduce the system to a one parameter family of equations, by letting $x\rightarrow x\sqrt{D/\gamma}$ and $t\rightarrow t/\gamma$. The hydrodynamic equations, obtained in a similar way to Theorem \ref{thm:hydro1} then rewrite, parametrized by the P\'eclet number $\Pe=\lambda/\sqrt{ D\gamma} $

\[\begin{split}
\partial_t \rho&=\Delta \rho-\Pe \partial_{u_1} m(1-\rho)\\
\partial_t m&=\Delta m-\Pe \partial_{u_1} \rho(1-\rho)-2 m
 \end{split}.\]

Thanks to this explicit macroscopic equation, using analogous techniques to the ones used in \cite{SSC18}, we perform in \cite{KEBT} the linear stability analysis of the uniform profile $\rho_0\equiv \rho$, $m\equiv 0$ in order to derive the exact phase diagram for our microscopic model for MIPS represented in Figure \ref{fig:MIPS}.

\subsection{A simple alignment model}
\label{sec:Flockmodel}

The second model considered in \cite{KEBT} is a one-dimensional \emph{zero-range process}, that is $\eta_x:=(\sigma_x^+, \sigma_x^-)\in \N^2$, where $\sigma^\pm_x$ represents the number of particles of type $\pm$ present at site $x$. Once again, there are three components to the dynamics:
\begin{enumerate}[i)]
\item a particle at site $x$ jumps to $x\pm1$ at rate $DN^2$.
\item A particle of type $\pm$ at site $x$ jumps at site $x\pm1$ at rate $\lambda N$.
\item A particle of type $\pm$ changes type (becomes $\mp$) at rate 
\[c^\pm_x(\sigma_x^+, \sigma_x^-)=e^{\mp\beta(\sigma_x^+- \sigma_x^-)}.\]
\end{enumerate}

Note that there is no longer any exclusion rule, and that the only interactions between particles come from the flipping dynamics: aside from it, particles behave as independent random walkers.
Since the model is not an exclusion process, the two initial profiles $\rho_0^+$, $\rho_0^-:[0,1]\to [0,+\infty[$ are not bounded by $1$. We consider an initial state $\eta(0) $ for the system defined by the product Poisson measure
\[\Prob(\sigma^\pm_x(0)=k)=\frac{(\rho_0^\pm(x/N))^k}{k!}e^{-\rho_0^\pm(x/N)}.\]
Although this assumption on the initial distribution can be relaxed, this choice of measure is not fortuitous, since Poisson measures are equilibrium measures for the symmetric zero-range dynamics defined by rule i) (see Section \ref{sec:LE}). Once again, the macroscopic evolution of the system is characterized by the respective density fields for $\pm$ particles, which can equivalently be accessed, as in the previous section,  either by taking the expected value or by coarse-graining the configuration as in \eqref{eq:EDF} and \eqref{eq:CGF}.

\medskip

\begin{figure}
\includegraphics[width=10cm]{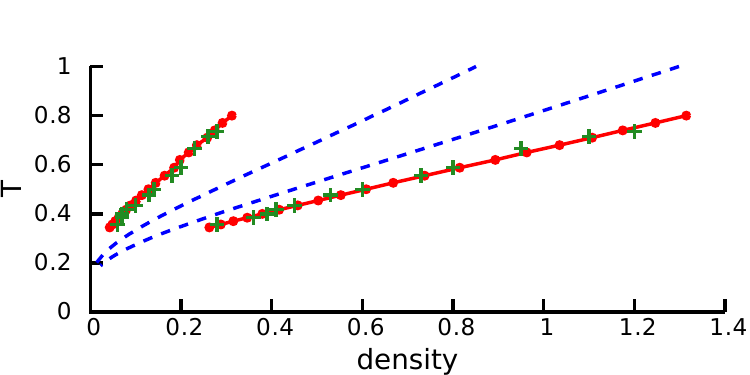}
\caption{From \cite{KEBT}: the phase diagram for the linear stability of the flocking model presented in Section \ref{sec:Flockmodel} in the $T=1/\beta$, $\rho_0\equiv\rho$ plane. The spinodal curves are represented in blue. Below the bottom spinodal curve, the "gaseous" phase $\rho\equiv \rho_0$, $m\equiv 0$ is linearly stable. Above the upper spinodal curve, the "liquid"  phase $\rho\equiv \rho_0$, $m\equiv m_0\neq 0$ is linearly stable. Between the two spinodals, a phase separated regime is observed, in which a magnetized ($\rho=\rho_\ell$, $m\neq 0$) band travels in a lower density gaseous ($\rho=\rho_g$, $m=0$) phase. The coexisting densities $\rho_g$ and $\rho_\ell$ are then given by the binodal curves in red.  Comparison with simulations of the microscopic system (green crosses) once again show perfect agreement with the analytical densities.}
\label{fig:flock}
\end{figure}

For this alignment model, still as a consequence of the  \emph{entropy method}, we have the following result.
\begin{theorem}
\label{thm:hydro2}
The macroscopic density and magnetization fields $\rho=\rho^++\rho^-$ and $m=\rho^+-\rho^-$ are solution to the coupled equations 
\[\begin{split}
\partial_t \rho&=D\partial_{uu} \rho-\lambda \partial_u m\\
\partial_t m&=D\partial_{uu} m-\lambda \partial_u \rho-2F(m),
 \end{split}\]
with the initial condition $\rho(0,\cdot)=\rho_0^++\rho_0^-$, $m(0,\cdot)=\rho_0^+-\rho_0^-$, where $F(m)$ is the explicit function 
\[F(\rho,m)=\pa{m\cosh[m\sinh(\beta)]-\rho\sinh[m\sinh(\beta)]}e^{-\beta+\rho\cosh(\beta)-\rho}.\]
\end{theorem}
	Given the alignment Glauber dynamics, this model displays a similar alignment phase transition as Vicsek's model. Once again, the derivation of the hydrodynamic equations allow us to perform an explicit linear stability analysis to obtain the exact phase diagram for the emergence of traveling bands, represented in Figure \ref{fig:flock}.

\subsection{Local equilibrium for active processes}
\label{sec:LE}

Local equilibrium is a key concept in the mathematical theory of hydrodynamic limits. Roughly speaking, a given microscopic system with a number of locally conserved quantities is said to satisfy local equilibrium if its local distribution around site $x$ is well approximated by the equilibrium distribution of the underlying microscopic Markov dynamics, parametrized by the local coarse-grained conserved quantities.

\medskip

 More explicitly, consider for example the \emph{Symmetric Simple Exclusion Process}, or  SSEP, in which only one particle can occupy each site of the system, and where a particle jumps to an empty neighbor at constant rate $N^2$. One easily checks that Bernoulli product measures $\mu_\rho$ with fixed parameter $\rho\in [0,1]$ (where each site is independently occupied w.p. $\rho$) are reversible, in the sense that they satisfy \emph{detailed balance}, w.r.t. the SSEP dynamics. The SSEP starting from a given density profile $\rho_0$ is in a state of local equilibrium, if at a time $t$, and around site $x$, the distribution of the SSEP is close to $\mu_{\rho^N_x(t)}$: for any function $f$ depending on sites (microscopically) close to $x$, 
\[\E(f(\eta))\equiv \E_{\mu_{\rho^N_x(t)}}(f(\eta))+o_N(1),\]
where the left expectation is taken w.r.t. the true measure of the process, whereas the right-hand side is the expectation with respect to the equilibrium measure $\mu_\rho$, whose parameter $\rho=\rho^N_x(t)=\E(\eta_x(t))$ is the local value of the density field in the considered region. By virtue of the law of large numbers, local equilibrium is the reason why, in particular,  the expected density field (cf.\eqref{eq:EDF}) and the coarse-grained field (cf.\eqref{eq:CGF}) are asymptotically equivalent.

\medskip

In the case of more general microscopic  dynamics like the dynamics above with several components, as is the case for the active matter models presented here, the equilibrium states of the process are in general no  longer explicit due to the interplay between dynamics (e.g. jump dynamics and flipping dynamics). However, because of the scale separation between different parts of the dynamics (cf. Section \ref{sec:WAEPZR}), the symmetric jumps occur much more frequently, and the symmetric jump dynamics thus enforces its local equilibrium to the other parts of the dynamics. Assuming one can show that local equilibrium is conserved throughout the dynamics, the corresponding equilibrium measures are therefore those of the symmetric part of the dynamics. 

\medskip

In the case of the first model presented above in Section \ref{sec:MIPSmodel}, the stirring dynamics i) is reversible w.r.t. product measures $\mu_{\rho^+, \rho^-}$ parametrized by the respective densities of $\pm$ particles, $\rho^\pm\in [0,1]$. Their marginals are given by 
\begin{equation}
\label{eq:mu}
\begin{cases}
\mu_{\rho^+, \rho^-}(\eta_x=1)=\rho^+\\
\mu_{\rho^+, \rho^-}(\eta_x=-1)=\rho^-\\
\mu_{\rho^+, \rho^-}(\eta_x=0)=1-\rho^+-\rho^-
\end{cases}.
\end{equation}

\medskip

For the zero-range--type model presented in Section \ref{sec:Flockmodel}, however, the exclusion rule is no longer enforced, and the equilibrium distribution of the symmetric part of the dynamics are also given by product measures $\nu_{\rho^+, \rho^-}$, parametrized this time by unbounded densities $\rho^\pm>0$. Their marginals at each site $x$ are given by the Poisson measures
\begin{equation}
\label{eq:nu}
\begin{cases}
\nu_{\rho^+, \rho^-}(\sigma_x^+=n^+, \;\sigma_x^-=n^-)=\frac{(\rho^+)^{n^+}(\rho^-)^{n^-}}{n^+!n^-!}e^{-\rho^+-\rho^-}\\
\end{cases}.
\end{equation}

\medskip

Note that in both cases, the equilibrium distribution of the process can be built by considering the equilibrium distribution of the type-blind process (Bernoulli product measure for the SSEP, Poisson product measure for the zero-range independent random walks), and then assigning independently to each particle the type $\pm$ with respective probability $\rho^\pm/(\rho^++\rho_-)$. Because both of the processes above maintain a state of local equilibrium, the distribution of each process in any microscopic box is well-approximated by $\mu_{\rho^+, \rho^-}$ (first model) or $\nu_{\rho^+, \rho^-}$ (second model), $\rho^+$, $\rho^-$ being the coarse-grained densities in the microscopic box considered.

\subsection{Computation of the hydrodynamic limit}
\label{eq:computehydro}
We now comment on the computation of the hydrodynamic limit starting from the microscopic dynamics considered. We will quite naturally take as examples the two models introduced above, however the steps described here are very general and can be applied broadly, assuming that a state of local equilibrium is maintained throughout the evolution of the process. Since our microscopic  dynamics are defined by their effect on $\sigma^\pm$ rather than on their $\rho$ and $m$ counterparts $\sigma^++\sigma^-$ and $\sigma^+-\sigma^-$, we naturally obtain the equations $\rho^\pm$ rather than those on $\rho$ and $m$. Of course, once the evolution equations for $ \rho^\pm$ have been obtained, those for $\rho$ and $m$ follow straightforwardly by addition and subtraction. We will show how to obtain the hydrodynamic equation for $\rho^+$, the analogous equation for $\rho^-$ is obtained in the same way. 

\medskip

Since we are working with a discrete space, it is natural to consider weak solutions to the hydrodynamic limit, to smooth out the jumps of the configuration's evolution. Fix then a smooth test function $H:[0,1]\to\R$, and consider the discrete integral 
\[\frac{1}{N}\sum_{x=1}^N H(x/N)\sigma_x^+(t)\]
of the process $\sigma^+$ against $H$.

\medskip

Assuming that local equilibrium remains in force throughout the dynamics, and since $H$ is a smooth function, this discrete integral is asymptotically equal to $\int_0^1 H(u) \rho^+(t,u) du$, $\rho^+(t,u)$ being the expected (or, equivalently, coarse-grained) density field defined in \eqref{eq:EDF}-\eqref{eq:CGF}. Now remains to characterize the evolution of $\rho^+$. Given the microscopic dynamics, we can write, as a consequence of \emph{Dynkin's formula} (see Remark \ref{rem:Dynkin} below),
\begin{multline}
\label{eq:Dynkin}
\frac{1}{N}\sum_{x=1}^N H(x/N)\sigma_x^+(t)=\frac{1}{N}\sum_{x=1}^N H(x/N)\sigma_x^+(0)\\
+\int_{0}^t \frac{1}{N}\sum_{x=1}^NH(x/N)\cro{j^+_{x-1,x}(s)-j^+_{x,x+1}(s)+h^+_x(s)}ds +\;\mbox{\emph{fluctuations}}.
\end{multline}
In the identity above, 
\[j^+_{x,x+1}=DN^2j^{+,s}_{x,x+1}+\lambda N j^{+,a}_{x,x+1}\] 
is the total (instantaneous) current of particles of type $+$ along the edge $(x,x+1)$, with a component of order $N^2$ due to symmetric jumps, and a component of order $N$ due to asymmetric jumps. The functions $j^{+,s}_{x,x+1}$  and $j^{+,a}_{x,x+1}$ encompass the jump rates and constraints due to the type of dynamics considered (exclusion, zero-range, etc.). The second contribution above,  $h^+_x$, is function of the flipping rates of the dynamics. The fluctuation term can be computed explicitly, and is shown to vanish as $N\to\infty$.

\medskip

For both models presented above, we can write, recalling notation \eqref{eq:sigma}
\begin{equation}
\label{eq:symcur}
j^{+,s}_{x,x+1}=\sigma^+_{x}-\sigma_{x+1}^+, 
\end{equation}
whereas 
\[j^{+,a}_{x,x+1}=\sigma^+_{x}(1-\sigma_x), \quad \quad h^+_x(s)=\gamma(\sigma^-_x-\sigma^+_x)\]
for the first model (notice the factor $(1-\sigma_x)$ coming from the exclusion rule) and 
\begin{equation}
\label{eq:hx}
j^{+,a}_{x,x+1}=\sigma^+_{x}, \quad \quad  h^+_x=\sigma^-_xc^-_x(\sigma_x^+, \sigma_x^-)-\sigma^+_xc^+_x(\sigma_x^+, \sigma_x^-)
\end{equation}
for the second model. All these functions are local functions of the configuration. By performing discrete integrations by parts in \eqref{eq:Dynkin}, one transfers the discrete derivatives on the smooth function $H$, thus balancing out all the extra factors $N$: integrating by parts in \eqref{eq:Dynkin} the difference of currents absorbs a first factor $N$, whereas the second factor coming from the diffusive parts gets absorbed by integrating by parts the second gradient in \eqref{eq:symcur}.

\medskip

Once these integrations by parts are performed, deriving the hydrodynamic limit just requires being able to replace local functions (such as $j^a_{x,x+1}$, $h_x^+$) of the configuration by their expected value. First, because once again the test function $H$ is smooth, one can replace any local function $g_x$ in \eqref{eq:Dynkin} by its average over a coarse-grained box of size $N^\delta$, for $\delta\in [0,1)$ 
\[\langle g\rangle ^{N^\delta}_x:=\frac{1}{2N^\delta+1}\sum_{|y-x|\leq N^\delta} g_y.\]
Because of \emph{local equilibrium} (cf Section \ref{sec:LE}), and by virtue of the law of large numbers, this quantity is asymptotically equal to $ \E_{\rho_+,\rho_-}(g_x)$, where the expectation is taken w.r.t. the relevant equilibrium measure ($\mu_{\rho_+, \rho_-}$ defined in \eqref{eq:mu} for the first model, $\nu_{\rho_+, \rho_-}$ defined in \eqref{eq:nu} for the second). Those measures are parametrized by the coarse-grained densities 
\[\rho ^{ N^\delta}_\pm(t,x):=\frac{1}{2N^\delta+1}\sum_{|y-x|\leq N^\delta} \sigma^\pm_y \underset{N\to\infty, \; x=\lfloor uN\rfloor}{\longrightarrow}\rho^{\pm}(t,u).\]

Deriving the hydrodynamic limit is then, at this point, merely a question of computing explicit expected values of local functions w.r.t. explicit equilibrium distributions. As an example, the function $F(\rho,m)$ appearing in Theorem \ref{thm:hydro2} can be written as the expectation 
\[F(\rho,m)=\widetilde{F}(\rho^+,\rho^-)=2\E_{\nu_{\rho^+, \rho^-}}(h_0^+),\]
where $h_0^+$ is the microscopic creation rate of $+$ particles, defined in \eqref{eq:hx}, and the factor $2$ comes from the fact that when a spin flips, the respective contribution to the magnetization is twice the variation of $\rho^+$. Further note that the function $F(\rho,m)$ is different from the mean-field magnetization rate.

\medskip

The steps presented above can be used fairly generally, and provide a robust road-map to derive hydrodynamic limits. Its main ingredients are:
\begin{enumerate}[i)]
\item determination of the equilibrium distributions of the dominant part of the dynamics (here, the diffusive one).
\item Showing that local equilibrium holds.
\item Thanks to the dynamics's jump rates, identifying the local functions (e.g. $j_{x,x+1}$, $h_x$) characterizing the microscopic dynamics.
\item Computing their averages under the equilibrium distribution.
\end{enumerate}

Of course, most of the mathematical difficulty to accomplish this program resides in proving ii), which can require a lot of work depending on the specificities of each model considered.

\begin{remark}[Dynkin's formula and fluctuation estimation]
\label{rem:Dynkin}
Let us comment briefly on Dynkin's formula, used to obtain  identity \eqref{eq:Dynkin}. For more details on the topic, we refer the interested reader to e.g. \cite[Appendix 1, Sections 3-5, pages 321-331]{KL}. Dynkin's formula holds for any continuous-time general Markov processes $( X_t)_{t\geq 0}$, and yields that for any bounded function $F$, 
\[M_t^F:=F(X_t)-F(X_0)+\int_{0}^t L F(X_s)ds\]
is a  \emph{martingale}. The operator $L$ is the \emph{generator of the process}; in the case of a pure jump process like the ones considered in this note, for example, assuming that $X$ jumps at rate $c(x\to x')$ from state $x$ to state $x'$, the generator writes 
\[LF(x)= \sum_{x'}c(x\to x')[F(x')-F(x)].\]
The $L^2$ norm of this martingale can be explicitly computed, 
\[\E\cro{(M_t^F)^2}=\E\cro{\int_0^t \sum_{x'}c(X_s\to x')[F(x')-F(X_s)]^2ds}.\]
Applying this identity to our lattice gases, we obtain that the $L^2$ norm $\E[(M_t^H)^2] $ of the fluctuations term in \eqref{eq:Dynkin} is 
\[E\cro{\int_0^t \frac{1}{N^2} \sum_{x=1}^N\pa{c(x,x+ 1, \eta(s))[\underset{=O(1/N)}{\underbrace{H(x'/N)-H(x/N)}}]^2+c(x, \eta(s))H(x/N)^2}ds}.\]
The rate $c(x,x+1, \eta)$ is the rate at which a particle jumps from $x$ to $x+1$ or from $x+1$ to $x$ in $\eta$ (as a result of  either symmetric or weakly asymmetric jumps) and is of order $N^2$, whereas $c(x,\eta)$ is the rate at which a particle at $x$ is flipped in $\eta$ and is of order $1$. The expectation above is therefore of order $O(1/N)$ and vanishes in the limit $N\to\infty$. 

\end{remark}

\subsection{Fluctuating hydrodynamics}

Equilibrium fluctuations for lattice gases is a fairly well understood topic, and some general tools allow for the explicit derivation of the equilibrium fluctuation field around its hydrostatic limit (see e.g. \cite[Chapter 11]{KL} for a look a general approach and specific references). As for the hydrodynamic limit, such tools rely strongly on the explicit knowledge of the equilibrium distribution of the underlying dynamics. 

\medskip

Non-equilibrium fluctuations, on the other hand, are still only partially understood from a mathematical standpoint. Some progress has been achieved making use of properties like duality or integrability, however such properties are strongly model-dependent and do not provide robust general methods to derive non-equilibrium fluctuations. Significant progress has been achieved on this front in \cite{JM}, where the non-equilibrium fluctuations for the Weakly Asymmetric Simple Exclusion Process (WASEP) are obtained in dimension $d\leq 3$, adapting Yau's  so-called \emph{relative entropy method} \cite{Yau1991} together with refined entropy estimates. 

\medskip

Consider the \emph{fluctuation fields} for the  MIPS model described in section \ref{sec:MIPSmodel}, defined by
\[\begin{split}
\mathcal{R}_t^{N}(H):=& \frac{1}{\sqrt{N}}\sum_{x=1}^N H(x/N)\pa{|\eta_x(t)|-\rho(t,x/N)}\\
\mathcal{M}_t^{N}(H):=& \frac{1}{ \sqrt{N}}\sum_{x=1}^N H(x/N)\pa{\eta_x(t)-m(t,x/N)},\\
\end{split}
\]
where the hydrodynamic fields $\rho$ and $m$ are given by Theorem \ref{thm:hydro1}. Although \cite{JM} is concerned with the exclusion process with one particle type without particle creation and destruction, one can expect that the arguments laid out in \cite{JM} could be adapted to the setting of the MIPS model presented above. One would therefore expect that  $(\mathcal{R}_t^{N},\mathcal{M}_t^{N})$ converges in a weak sense to limiting fields $(\mathcal{R}_t,\mathcal{M}_t),$ which are solution to the coupled  SPDEs
\begin{equation}
\label{eq:flHDL}
\begin{split}
\partial_t \mathcal{R}&=D \Delta \mathcal{R}- 2 \lambda \partial_{u_1}[(1-\rho)\mathcal{M}-m\mathcal{R}]+\dot{\mathscr{W}}^{\mathcal{R}}_t \\
\partial_t \mathcal{M}&=D \Delta \mathcal{M}- 2 \lambda \partial_{u_1}(1-2\rho)\mathcal{R} -2 \gamma \mathcal{M}+  \dot{\mathscr{W}}^{\mathcal{M}}_t +\sqrt{2 \gamma}\dot{\mathscr{B}}_t,
 \end{split}
\end{equation}
where 
\[
\begin{split}
\dot{\mathscr{W}}^{\mathcal{R}}_t=\sqrt{2D\rho^+(1-\rho^+)}\dot{\mathscr{W}}^+_t +\sqrt{2D\rho^-(1-\rho^-)}\dot{\mathscr{W}}^-_t\\
\dot{\mathscr{W}}^{\mathcal{M}}_t=\sqrt{2D\rho^+(1-\rho^+)}\dot{\mathscr{W}}^+_t -\sqrt{2D\rho^-(1-\rho^-)}\dot{\mathscr{W}}^-_t,
\end{split}
\]
$\dot{\mathscr{W}}^\pm_t$, are two independent space-time white noises, and $\dot{B}_t$ is a white noise in time independent from  $\dot{\mathscr{W}}^\pm$. The first line of equation \eqref{eq:flHDL} comes from the Taylor expansion of the function $Q(\rho, m)=m(1-\rho)$ (appearing  in the ballistic term of the hydrodynamic limit) in its two variables, in the sense that 
\[Q(\rho', m')-Q(\rho,m)= (1-\rho)(m'-m)-m(\rho'-\rho)+error,\]
the error term being of a higher order in $\max (\rho-\rho', m-m')$ and  therefore vanishes in the limit. This yields the "finite but large" formal fluctuating hydrodynamics for this model given by
\begin{equation*}
\begin{split}
\partial_t \rho&=D\partial_{uu} \rho-\lambda \partial_u m(1-\rho)+\frac{1}{\sqrt{N}}\dot{\mathscr{W}}^{\mathcal{R}}_t,\\
\partial_t m&=D\partial_{uu} m-\lambda \partial_u \rho(1-\rho)-2\gamma m +\frac{1}{\sqrt{N}} \cro{ \dot{\mathscr{W}}^{\mathcal{M}}_t +\sqrt{2 \gamma}\dot{\mathscr{B}}_t}.
 \end{split}
\end{equation*}
Let us emphasize once again that this formal statement is at this point merely a conjecture, and that technical difficulties would most likely need to be overcome to adapt arguments in \cite{JM} to this setting. For this reason, we do not write here a formal equation for the macroscopic fluctuations for the second model presented in Section \ref{sec:Flockmodel}, since it is not likely to be a (somewhat) straightforward consequence of  \cite{JM}.

\section{A look at non-gradient active lattice gases}
\label{sec:NG}

\subsection{Non-gradient hydrodynamics}

The scheme of proof presented above has been used with great success on a number of--at least in parts--diffusive microscopic models since its invention by Guo, Papanicolaou and Varadhan. However, it strongly relies on the discrete integration by parts in order to balance out the $N$ factors coming from the time rescaling of the jump dynamics. One of those integrations by parts is always available, since it comes from the difference of currents coming in and currents going out of a given site (see equation \ref{eq:Dynkin}). 

\medskip

The second integration by parts, however, which allows to balance out the extra factor $N$ of the diffusive dynamics, requires the symmetric current along an edge $(x,x+1)$ to be of the form \eqref{eq:symcur}, or at least of the form 
\begin{equation}
\label{eq:gradient}
j^s_{x,x+1}=g_x(\sigma)-g_{x+1}(\sigma)
\end{equation} 
for some local function $g_x$ of the configuration around $x$. Models satisfying \eqref{eq:gradient} are called \emph{gradient}, and usually allow for explicit derivation of their hydrodynamic limit. Many, if not most, models, however, do not satisfy the gradient condition, and this strongly limits one's ability to derive hydrodynamic limits, and scaling limits of lattice gases in general.

\medskip

This issue was overcome by Varadhan \cite{Varadhan1994b}, who derived general tools to obtain hydrodynamic limits for non-gradient models. However, his technique, although widely applicable, still suffers from significant shortcomings. The main drawback is the necessity for a sharp estimate of the spectral gap of the diffusive dynamics, ensuring that one can obtain $\mathscr{L}^2$ estimates on local functions, by the so-called Dirichlet form 
\[\E(g^2)\leq CN^2\mathscr{D}(g).\]
Such estimates are crucial for the non-gradient method, since they allow to deduce from the estimation of the Dirichlet form $\mathscr{D}(g)$, obtained from the dynamics itself, direct estimates on the system's observables. 

\medskip

In theory, spectral gap estimates can be obtained by slightly relaxing the dynamics considered, by adding  for example some stirring dynamics, (i.e. dynamics i) of Section \ref{sec:MIPSmodel}), at a small rate $N^{\delta} \ll N^2$ with no explicit impact on the hydrodynamic limit itself.  However, Varadhan's non-gradient method does not work well with such relaxed dynamics, since it relies on infinite volume estimates which do not allow for perturbative arguments. This requirement for a sharp estimate on the spectral gap is the first main issue with the non-gradient method.

\bigskip

A second obstacle in the study of non-gradient systems is the significant technical difficulty of Varadhan's non-gradient estimates, as well as the relative obscurity of the inner workings of his method. Even today, it is still a rather inaccessible piece of mathematics, often used as a black box by a large part of the mathematical community concerned with scaling limits of interacting particle systems.

\bigskip

The third shortcoming of the non-gradient method is also the most important in the context of this note, and is a consequence of the non-constructive nature of the method. The aim of Varadhan's non-gradient method is to show a so-called \emph{microscopic fluctuation-dissipation equation} 
\begin{equation}
\label{eq:gradient2}
j^s_{x,x+1}\simeq D(\rho_x)(\sigma_{x+1}-\sigma_x)+\mathscr{L}f,
\end{equation} 
where the second term $ \mathscr{L}f$ is a small fluctuation that disappears at the hydrodynamic limit. A significant drawback is that the effective \emph{diffusion coefficient} $D$ is not an explicit function of the density, and is instead obtained via a variational formula. Roughly speaking, this is due to a deformation of local equilibrium due to the non-gradient nature of the dynamics, which distorts in a non-explicit way the local distribution of the process, thus preventing from writing the macroscopic observables of the system as explicit expected values  of the microscopic ones. Similarly, in the case of weakly asymmetric models where particles are subject to weak driving forces, the ballistic term $G(\rho)$ appearing in the hydrodynamic limit 
\[\partial_t\rho=\partial_u D(\rho) \partial_u \rho+\partial_u G(\rho)\]
is also non-explicit, although it satisfies Einstein's relation
\[D(\rho)=\chi(\rho)G(\rho),\]
$\chi(\rho)$ denoting the system's compressibility.

\medskip

Naturally, this lack of explicit  formulas for the hydrodynamic limit's coefficients is a significant setback in the context of active lattice gases, where one's goal in deriving the hydrodynamic limit is to obtain some phenomenological understanding of the macroscopic behavior of the system.

\subsection{The active exclusion process}
\label{sec:nongradient}

To emphasize how easily one can stumble upon non-gradient models, we slightly modify the MIPS model described in Section \ref{sec:MIPSmodel}. For reasons that will be discussed later on consider the \emph{two-dimensional model}, that we will call  \emph{Active Exclusion Process} (AEP), with the following dynamics:
\begin{enumerate}[i)]
\item a particle at site $x$ jumps to any neighboring \emph{empty} site $x\pm e_i$ at rate $DN^2$.
\item A particle $\pm$ at site $x$ jumps at site $x\pm e_1$ at rate $\lambda N$ \emph{if it is empty}.
\item A particle changes type at constant rate $\gamma$.
\end{enumerate}

Note that aside from the dimension (which adds no extra difficulty to the derivation of the hydrodynamic limit of the MIPS model presented in \ref{sec:MIPSmodel}), the only difference between the two models is that in the AEP, the \emph{exclusion rule} is also enforced on the symmetric part of the dynamics. Since swapping two identical particles has no effect on the system, the only dynamical difference is therefore the impossibility  for two neighboring particles with different types $+\;-$ to switch positions.

\medskip

In trying to apply the scheme of proof presented in Section \ref{eq:computehydro}, one obtains, because of the exclusion rule, that the symmetric $+$ particle current along edge $(x, x+e_i)$ is this time  given by 
\begin{equation*}
j^{+,s}_{x,x+e_i}=\sigma^+_{x}(1-\sigma_{x+e_i})-\sigma_{x+e_i}^+(1-\sigma_{x}), 
\end{equation*}
which cannot be expressed as a discrete gradient $h_{x+e_i}-h_x$.
Note that this is not a problem for type-blind exclusion, for which 
\begin{equation*}
j^{s}_{x,x+e_i}=\sigma_{x}(1-\sigma_{x+e_i})-\sigma_{x+e_i}(1-\sigma_{x})=\sigma_{x}-\sigma_{x+e_i}
\end{equation*}
is a discrete gradient.

\medskip

The second integration by parts can't therefore be performed on the AEP, and one needs to use Varadhan's non-gradient method to obtain a  \emph{microscopic fluctuation-dissipation equation} 
\begin{equation*}
j^{+,s}_{x,x+1}\simeq d_s(\rho_x)(\sigma^+_{x+1}-\sigma^+_x)+D(\rho^+_x, \rho_x)(\sigma_{x+1}-\sigma_x)+\mathscr{L}f,
\end{equation*}
analogous to \eqref{eq:gradient2} and derive its hydrodynamic limit. 

\medskip

Note the two components to the symmetric current of particles : the first one, with diffusion coefficient  $d_s(\rho_x)$, quantifies the capacity of the SSEP to mix different types of particles in a homogeneous overall setting. Consider for example an initial macroscopic state with constant overall density $\rho_0\equiv c\in (0,1)$, but with segregated types,
\[ \rho_0^+(u)=c {\bf 1}_{[0,1/2]}(u), \quad  \rho_0^-(u)=c{ \bf 1}_{(1/2,1]}(u).\]
Assuming for example that the system is only affected by the diffusive part of the dynamics, under its influence, after a long time $t>>1$, such a segregated macroscopic profile should be able to relax to the uniform macroscopic state
\[\rho^+(t,u)=\rho^+(t,u)\simeq\frac c2\quad \forall u\in [0,1].\]
This is the effect of the contribution $d_s(\rho_x)(\sigma^+_{x+1}-\sigma^+_x)$ in \eqref{eq:gradient2}.

 The coefficient $d_s(\rho)$, called \emph{self-diffusion coefficient}, is also the diffusion coefficient of a tagged tracer particle in a homogeneous environment with density $\rho$, and vanishes as $\rho\to1$. This quantity is the reason why the model is discussed in two dimensions or more : in dimension $1$, $d_s(\rho)\equiv 0$, because particles have no room to go around each other, so that the hydrodynamic limit of the $1$-dimensional AEP is trivial. 
However, even in two dimensions, the low mixing at high densities is a significant issue in order to derive the hydrodynamic limit. 

\medskip

The second diffusion coefficient $D(\rho^+_x, \rho_x)$ is more straightforward, and quantifies the system's capacity to smooth out its total (in the sense of $\rho=\rho^++\rho^-$) density heterogeneities, what might be referred to as "standard" diffusion. Once again, the fluctuation $\mathscr{L}f$ disappears in the hydrodynamic scaling. 

\medskip

As a simpler case of \cite{Erignoux}, assuming that the microscopic model is in an initial state given by \eqref{eq:initstate}, one can state the following result.
\begin{theorem}
\label{thm:hydro3}
Assume that the initial particle density is bounded away from $1$, 
\[0\leq \rho_0^+(u)+\rho_0^-(u)<1\quad  \forall u\in [0,1].\]
Then, the macroscopic density fields $\rho^+$ and $\rho^-$ are solution to the coupled equations 
\[\begin{split}
\partial_t \rho^+&=\nabla \cro{d_s(\rho)\nabla \rho^++D(\rho^+, \rho) \nabla\rho}-\lambda \partial_{u_1}S(\rho^+, \rho)-\gamma m\\
\partial_t \rho^-&=\nabla \cro{d_s(\rho)\nabla \rho^-+D(\rho^-, \rho) \nabla\rho}+\lambda \partial_{u_1}S(\rho^-, \rho)+\gamma m\\
 \end{split}\]
where once again $\rho$ and $m$ are  respectively the total density and magnetization fields, with  the initial condition $\rho^\pm(0,\cdot)=\rho_0^\pm$.
\end{theorem}

Both $D$ and $S$ are (explicit) functions of $\rho^+$, $\rho^-$ and $d_s(\rho)$. Because we used Varadhan's non-gradient tools, however, $d_s(\rho)$ itself is not an explicit function of $\rho$, and is defined through a variational formula (see  \cite{Spohn1990}). As discussed in Section \ref{sec:continuum} below, this result is a simpler case of the model studied in \cite{Erignoux}, where particle types $\theta\in[0,2\pi[$ take their value in a continuum. For non-gradient models, no tools currently exist to derive non-equilibrium fluctuations, therefore deriving the fluctuating hydrodynamics for this type of models is completely out of reach.

\section{On the coarse-graining scale $N^\delta$ }
\label{ref:CGscale}

From a mathematical standpoint, although the definition \eqref{eq:CGF} of the coarse-grained fields as limiting macroscopic quantities of interest is correct, they are not in general the basis for the mathematical proof of hydrodynamic limits. Instead, the most classical tools to derive hydrodynamic limits rely on the widely used \emph{one-block} and \emph{two-blocks estimates}, which together with a compactness argument, are usually sufficient to derive the hydrodynamic limit. 

\medskip 

Consider for example a given lattice gas with the particle density as the only locally conserved quantity. 
The \emph{one-block estimate} concerns the microscopic scale, that is a scale $\ell$ going to $\infty$ \emph{after} the hydrodynamic scaling parameter $N$. More precisely, the one-block estimate states that a law of large numbers holds on asymptotically large \emph{microscopic} boxes, in the sense that, denoting by $\tau_y$ the discrete translation of a function by $y$, and given a local function of the configuration $g$,
\[\lim_{\ell\to\infty}\lim_{N\to\infty}\cro{\frac{1}{|B_\ell|}\sum_{x\in B_\ell}\tau_{x+y}g-\E_{\rho_x^\ell}(g)}=0.\]
Above,  the expectation is taken w.r.t. the equilibrium measure parametrized by the coarse-grained conserved quantities $\rho_x^\ell$ over the microscopic box $B_\ell(x)$ around site $x$. Note that the two limits above do not commute, so that $ \ell$ should be thought of as the "large but finite" size of a microscopic box in the infinite lattice $N\to\infty$. We skip for the sake of brevity and clarity the mathematical details of the one-block estimate, and refer to Lemma 3.1, p. 82 of \cite{KL} for a detailed implementation. The one-block estimate is fairly widely available, and is one of the main ingredients of most techniques deriving  scaling limits of interacting particle systems, be it on the scale of the law of large numbers (hydrodynamic limits), central limit theorem (fluctuations around the hydrodynamic limit), or large deviations principles.

\medskip

The \emph{two-blocks estimate} concerns mesoscopic scales, that is, scales of order $\varepsilon N$, with $\varepsilon$ going to $0$ \emph{after} the scaling parameter $N$ was sent to $\infty$.  The two-blocks estimate states that the microscopically coarse-grained field $\rho^\ell_x$, with $\ell$ going to infinity after both $N\to\infty$ and $\varepsilon \to 0$, does not vary on mesoscopic scales, that is 
\[\lim_{\ell\to\infty}\lim_{\varepsilon\to 0}\lim_{N\to\infty} \sup_{|y|\leq \varepsilon N}\cro{\rho_x^\ell - \rho_{x+y}^\ell}=0,\]
The two-blocks estimate, together with the one-block estimate, and the Lipschitz-continuity of $\E_\rho(g)$ in the parameter $\rho$, allows the replacement of microscopic spatial averages 
\[\frac{1}{|B_\ell|}\sum_{x\in B_\ell}\tau_{x+y}g\] 
(which can be introduced at no cost because of the test function, cf. Section \ref{eq:computehydro}) by their expectation w.r.t. to equilibrium measures whose parameter $\rho^{\varepsilon N}_x$ is coarse-grained on the \emph{mesoscopic scale} $\varepsilon N$. The big upside is that mesoscopic coarse-grained quantities remain tractable in the limit $N \to\infty$, so that $\E_{\rho^{\varepsilon N}_{\lfloor uN\rfloor}(t)}(g)$ admits a well-defined macroscopic limit $\E_{\rho^{(\varepsilon)}(t,u)}(g)$, whose limit $\varepsilon \to 0$ in turn converges to an explicit function $\E_{\rho(t,u)}(g)$ of the hydrodynamic limit.

\medskip

Unfortunately, although the \emph{one-block estimate} is a fairly generally available tool, the \emph{two-blocks estimate} is more fragile, and for many models it does not hold. For example, asymmetric (and not \emph{weakly} asymmetric models) and boundary-driven systems do not, except in very specific cases, allow for the use of the two-blocks estimate. In such cases, other techniques must be considered. This can sometimes lead to weaker forms of the hydrodynamic limit, for example as entropy solutions and/or measure-valued solutions to the hydrodynamic equation (see for example chapter 9 of \cite{KL} and references therein).

\medskip

Such models raise a number of questions on the limiting PDE's, mainly regarding the uniqueness of solutions (in general, the existence of the hydrodynamic limit yields existence of solutions), in order to discard the non-physically relevant ones. In most cases, this can be done by exploring the (sometimes dense) mathematical PDE literature, and is not a significant obstacle. In the context of active matter however, hydrodynamic limits can typically involve cross-diffusion equations with both ballistic and reaction terms, which are not always covered by the state of the art, and for which uniqueness of weak solutions is not guaranteed, either because a relevant energy estimate cannot be proved from the microscopic system, or because such an energy estimate is not sufficient to guarantee uniqueness. This was, for example, the case of the non-gradient model  studied in \cite{Erignoux} briefly described in Section \ref{sec:nongradient}. For this reason, the main result in \cite{Erignoux} proves the convergence of the microscopic model to a set of solutions of the hydrodynamic limit satisfying an energy estimate making the PDE well-defined, but does not prove that  the set is a singleton.

\section{generality of the results and extensions}
\label{sec:extensions}

We now mention some relevant extensions of the results presented above.

\subsection{extension to more particle types}
\label{sec:continuum}

The results presented above readily extend to more particle types : in general, although such extensions can complicate the analysis of the limiting PDE, they do not pose a significant issue to the mathematical derivation of the hydrodynamic limit. For "almost" linear models as the ones presented in Sections \ref{sec:MIPSmodel} and \ref{sec:Flockmodel}, even though the full macroscopic description might involve all particle types, one can obtain in some cases closed coupled equations for the density and magnetic fields, allowing for the derivation of phase diagrams as in \cite{KEBT}.

\medskip

The extension to a continuum of particle types, although it induces more technical difficulties, does not in general pose phenomenological hurdles to the hydrodynamic limit. As an example, as mentioned before, the model studied in \cite{Erignoux} concerns particles, each characterized by their velocity's angle in the plane, denoted by $\theta$,  performing weakly asymmetric random walks with weak drift $\lambda N \pa{\begin{matrix}\cos\theta\\ \sin \theta\end{matrix}}$. In this perspective, the two particle types $\pm$ correspond to the choices $\theta=0$ and $\theta=\pi$ respectively. In this case, although the functions $\cos$ and $\sin$ naturally make sense from a modeling standpoint, they can be replaced by any smooth function of the particle's parameter $\theta$ and still allow for the derivation of the hydrodynamic limit.

\subsection{impact of the dimension}

Any of the models briefly presented above can be defined in higher dimension. The only significant hurdle in dimension $1$ has already been pointed out and  concerns the non-gradient model introduced in Section \ref{sec:nongradient}, whose phenomenology is different in dimension $1$ where the self diffusion coefficient $ d_s(\rho)$ vanishes for every $ \rho$. As a result, the hydrodynamic limit for the $1$-dimensional non-gradient model does not hold due to the lack of local  mixing between $+$ and $-$ particles. The drift applied to the particles can be higher-dimensional as well to match the spatial dimension of the particle's motions.

\medskip

Note that we are concerned here only with the derivation of the hydrodynamic limit, not the analysis of its behavior with the objective of understanding the macroscopic behavior of the model:  the latter can be deeply impacted by the dimension change, even when it does not pose a problem from a hydrodynamic perspective.

\bibliographystyle{plain}
\bibliography{biblio}

\begin{thebibliography}{10}

\bibitem{DB14}
A.B.~T. {Barbaro} and P.~{Degond}.
\newblock {Phase transition and diffusion among socially interacting
  self-propelled agents}.
\newblock {\em Discrete and Continuum Dynamical Systems B,19 pp. 1249-1278},
  2014.

\bibitem{Alignment2}
A.~Baskaran and M.~C. Marchetti.
\newblock Self-regulation in self-propelled nematic fluids.
\newblock {\em The European Physical Journal E, Soft Matter}, page 35(9):95,
  September 2012.

\bibitem{rods}
M.~Bär, R.~Großmann, S.~Heidenreich, and F.~Peruani.
\newblock Self-propelled rods: Insights and perspectives for active matter.
\newblock {\em Annual Review of Condensed Matter Physics}, 11(1):441--466,
  2020.

\bibitem{CT2015}
M.~E. {Cates} and J.~{Tailleur}.
\newblock Motility-induced phase separation.
\newblock {\em Annual Review of Condensed Matter Physics}, 6:219--244, March
  2015.

\bibitem{Alignment0}
H.~Chat\'e, F.~Ginelli, G.~Gr\'egoire, and F.~Raynaud.
\newblock Collective motion of self-propelled particles interacting without
  cohesion.
\newblock {\em Phys. Rev. E}, 77:046113, Apr 2008.

\bibitem{DM2007}
P.~{Degond} and S.~{Motsch}.
\newblock Continuum limit of self-driven particles with orientation
  interaction.
\newblock {\em Mathematical Models and Methods in Applied Sciences}, 18,
  October 2008.

\bibitem{DY2010}
P.~Degond and T.~Yang.
\newblock Diffusion in a continuum model of self-propelled particles with
  alignment interaction.
\newblock {\em Mathematical Models and Methods in Applied Sciences},
  20:1459--1490, February 2010.

\bibitem{Erignoux}
C.~Erignoux.
\newblock Hydrodynamic limit for an active exclusion process.
\newblock {\em M\'emoires de la Soci\'et\'e Math\'ematique de France}, (169),
  2021.

\bibitem{Frouvelle2011}
A.~{Frouvelle}.
\newblock A continuum model for alignment of self-propelled particles with
  anisotropy and density-dependent parameters.
\newblock {\em Mathematical Models and Methods in Applied Sciences}, 22(7),
  December 2012.

\bibitem{FL2012}
A.~Frouvelle and J.-G. Liu.
\newblock Dynamics in a kinetic model of oriented particles with phase
  transition.
\newblock {\em SIAM Journal on Mathematical Analysis}, 44(2):791--826, 2012.

\bibitem{animal}
I.~Giardina.
\newblock Collective behavior in animal groups: theoretical models and
  empirical studies.
\newblock {\em HFSP journal}, 2(4):205—219, August 2008.

\bibitem{GC2004}
G.~{Gr{\'e}goire} and H.~{Chat{\'e}}.
\newblock Onset of collective and cohesive motion.
\newblock {\em Physical Review Letters}, 92(2):025702, January 2004.

\bibitem{GPV1988}
M.Z. {Guo}, G.C. {Papanicolaou}, and S.R.S. {Varadhan}.
\newblock Nonlinear diffusion limit for a system with nearest neighbor
  interactions.
\newblock {\em Commun. Math. Phys.}, 118(1):31--59, 1988.

\bibitem{Alignment1}
T.~Ihle.
\newblock Kinetic theory of flocking: Derivation of hydrodynamic equations.
\newblock {\em Phys. Rev. E}, 83:030901, Mar 2011.

\bibitem{JM}
Milton {Jara} and Ot{\'a}vio {Menezes}.
\newblock {Non-equilibrium Fluctuations of Interacting Particle Systems}.
\newblock {\em arXiv e-prints}, page arXiv:1810.09526, October 2018.

\bibitem{KL}
C.~Kipnis and C.~Landim.
\newblock {\em Scaling limits of interacting particle systems}, volume 320 of
  {\em Grundlehren der Mathematischen Wissenschaften [Fundamental Principles of
  Mathematical Sciences]}.
\newblock Springer-Verlag, Berlin, 1999.

\bibitem{KEBT}
M.~{Kourbane-Houssene}, C.~{Erignoux}, T.~{Bodineau}, and J.~{Tailleur}.
\newblock {Exact Hydrodynamic Description of Active Lattice Gases}.
\newblock {\em Physical Review Letters}, 120(26):268003, June 2018.

\bibitem{MCN20}
D.~{Martin}, H.~{Chat{\'e}}, C.~{Nardini}, A.~{Solon}, J.~{Tailleur}, and
  F.~{van Wijland}.
\newblock {Fluctuation-induced phase separation in metric and topological
  models of collective motion}.
\newblock {\em arXiv e-prints}, page arXiv:2008.01397, August 2020.

\bibitem{SFC20}
X.-Q. Shi, G.~Fausti, H.~Chat\'e, C.~Nardini, and A.~Solon.
\newblock Self-organized critical coexistence phase in repulsive active
  particles.
\newblock {\em Phys. Rev. Lett.}, 125:168001, Oct 2020.

\bibitem{bacteria}
A.~Sokolov and I.~S. Aranson.
\newblock Physical properties of collective motion in suspensions of bacteria.
\newblock {\em Physical Review Letters}, 109:248109, Dec 2012.

\bibitem{SSC18}
A.~Solon, J.~Stenhammar, M.~E. Cates, Y.~Kafri, and J.~Tailleur.
\newblock Generalized thermodynamics of phase equilibria in scalar active
  matter.
\newblock {\em Phys. Rev. E}, 97:020602, Feb 2018.

\bibitem{Alignment3}
A.~P. Solon and J.~Tailleur.
\newblock Flocking with discrete symmetry: The two-dimensional active ising
  model.
\newblock {\em Phys. Rev. E}, 92:042119, Oct 2015.

\bibitem{Spohn1990}
H.~Spohn.
\newblock Tracer diffusion in lattice gases.
\newblock {\em Journal of Statistical Physics}, 59(5-6):1227--1239, 1990.

\bibitem{Varadhan1994b}
S.~R.~S. Varadhan.
\newblock non-linear diffusion limit for a system with nearest-neighbor
  interactions ii.
\newblock In {\em Asymptotic problems in probability theory : stochastic models
  and diffusion on fractals}, number 283 in Pitman Research Notes in
  Mathematics, pages 75--128. Springer-Verlag, 1994.

\bibitem{Vicsek}
T.~{Vicsek}, A.~{Czir{\'o}k}, E.~{Ben-Jacob}, I.~{Cohen}, and O.~{Shochet}.
\newblock Novel type of phase transition in a system of self-driven particles.
\newblock {\em Physical Review Letters}, 75:1226--1229, August 1995.

\bibitem{Yau1991}
H-T. Yau.
\newblock Relative entropy and hydrodynamics of ginzburg-landau models.
\newblock {\em Letters in Mathematical Physics}, 22.1:63--80, 1991.

\end{thebibliography}

\end{document}